\documentclass[aps,pre,twocolumn,floatfix,groupedaddress,footinbib,notitlepage,showpacs,superscriptaddress,longbibliography]{revtex4-1}
\usepackage{times,bm,bbm,bbold,graphicx,graphics,amssymb,amsmath,amsthm,amsfonts,dsfont,hyperref,color}
\usepackage{bbold}

\newcommand{\ket}[1]{|#1\rangle}
\newcommand{\bra}[1]{\langle#1|}
\newcommand{\braket}[2]{\langle #1|#2\rangle}
\newcommand{\ketbra}[2]{|#1\rangle\langle #2|}

\newcommand{\beq}{\begin{equation}}
\newcommand{\eeq}{\end{equation}}

\newcommand{\ignore}[1]{}

\newcommand{\be}{\begin{equation}}
\newcommand{\ee}{\end{equation}}
\newcommand{\vlambda}{\underline{\lambda}}
\newcommand{\vs}{\underline{s}}

\newcommand{\vK}{\underline{K}}
\newcommand{\valpha}{\underline{\alpha}}

\definecolor{mygreen}{rgb}{0, 0.7, 0}

\let\oldsqrt\sqrt
\def\sqrt{\mathpalette\DHLhksqrt}
\def\DHLhksqrt#1#2{%
\setbox0=\hbox{$#1\oldsqrt{#2\,}$}\dimen0=\ht0
\advance\dimen0-0.2\ht0
\setbox2=\hbox{\vrule height\ht0 depth -\dimen0}%
{\box0\lower0.4pt\box2}}

\DeclareFontFamily{OT1}{pzc}{}
\DeclareFontShape{OT1}{pzc}{m}{it}%
              {<-> s * [1.25] pzcmi7t}{}
\DeclareMathAlphabet{\mathpzc}{OT1}{pzc}%
                                 {m}{it}

\begin{document}

\title{Symmetry-induced fluctuation relations in open quantum systems}


\author{Stefano Marcantoni}

\email{stefano.marcantoni@nottingham.ac.uk}
\affiliation{School of Physics \& Astronomy, University of Nottingham, Nottingham NG7 2RD, UK}
\affiliation{Centre for the Mathematics and Theoretical Physics of Quantum Non-equilibrium
Systems, University of Nottingham, Nottingham NG7 2RD, UK}

\author{Carlos P\'erez-Espigares}

\email{carlosperez@ugr.es}
\affiliation{Departamento de Electromagnetismo y F\'isica de la Materia, Universidad de Granada, Granada 18071, Spain}
\affiliation{Institute Carlos I for Theoretical and Computational Physics, Universidad de Granada, Granada 18071, Spain}

\author{Juan P. Garrahan}

\email{juan.garrahan@nottingham.ac.uk}
\affiliation{School of Physics \& Astronomy, University of Nottingham, Nottingham NG7 2RD, UK}
\affiliation{Centre for the Mathematics and Theoretical Physics of Quantum Non-equilibrium
Systems, University of Nottingham, Nottingham NG7 2RD, UK}

\begin{abstract}
We derive a general scheme to obtain quantum fluctuation relations for dynamical observables in open quantum systems. For concreteness we consider Markovian non-unitary dynamics that is unraveled in terms of quantum jump trajectories, and exploit techniques from the theory of large deviations like the tilted ensemble and the Doob transform. Our results here generalise to open quantum systems fluctuation relations previously obtained for classical Markovian systems, and add to the vast literature on fluctuation relations in the quantum domain, but without resorting to the standard two-point measurement scheme. We illustrate our findings with three examples in order to highlight and discuss the main features of our general result.
\end{abstract}

\maketitle

\section{Introduction}

The discovery of fluctuation relations that hold true arbitrarily far from equilibrium in the 1990s \cite{evans93a, gallavotti95a,gallavotti95b, kurchan98a, lebowitz99a, jarzynski97b, crooks99a, crooks00a} boosted a lot of successive work on the topic; for reviews see \cite{evans02a,sevick08a,Ritort2008,jarzynski11a, seifert12a, gallavotti20a}. While the first results were in the framework of classical physics, quantum fluctuation relations were also discovered quite soon after \cite{kurchan00a, tasaki00a}. In the quantum setting one usually studies the statistics of stochastic observables obtained through the so-called two-point-measurement protocol. Despite being initially restricted to closed quantum systems, many results were eventually discovered for open quantum settings \cite{campisi09a}, both in the case of unital dynamics \cite{albash13a, rastegin14a, goold15a, aurell15a} and for generic dynamics \cite{jarzynski04a, chetrite12a, manzano15a, manzanog18a, jaksic14a, ramezani18a, ramezani18b}; for reviews see \cite{esposito09a, campisi11a}. 

In describing open quantum systems, most of the work is devoted to the study of master equations in so-called Lindblad form \cite{lindblad76a,Gorini1976} which describes the dynamics of the average state.  A different perspective can be obtained by looking at the unraveling of a master equation in terms of quantum jump trajectories \cite{Belavkin1990,Dalibard1992}. In general, many different unravelings correspond to the same master equation and without any information about the environment, one cannot distinguish between them. Instead, when the environment is continuously monitored through detectors, a specific unraveling can acquire a physical meaning, because the number of jumps of the wavefunction is mapped into the number of clicks in a detector (for reviews see \cite{plenio98a,gardiner04a}). 
In this framework one can look at the so-called dynamical observables, related to the number of jumps in a particular realization of the dynamics, as customarily done in the context of continuous-time Markov chains \cite{garrahan09a, garrahan16a}. 

In this paper we build on the thermodynamic formalism for quantum jump trajectories introduced in \cite{garrahan10a} to provide a general framework for quantum fluctuation relations. In particular, we generalise a classical approach presented recently in \cite{marcantoni20a} to the quantum domain. This approach allows us to formulate fluctuation relations in the statistics of dynamical observables given transformations in the space of trajectories for suitable observable and perturbed (biased) dynamics. Our result is obtained using techniques of large deviation theory 
\cite{touchette09a} like biased trajectory ensembles 
\cite{garrahan09a,garrahan16a,chetrite15b,manzano14a} and Doob transforms \cite{Borkar2003,jack10a,chetrite15b,carollo18b}.

The paper is structured as follows. In Section \ref{sec:traj} we review the formalism of quantum jump trajectories and consider the statistics of dynamical observables. In particular, we recall how in the the large deviation regime, the statistics are encoded in the properties of a suitable tilted generator, whose largest eigenvalue contains all the information about the long-time fluctuations. 
We also review the Doob transformation which allows us to obtain the quantum dynamics of the subset of trajectories leading to a given fluctuation. In Section \ref{sec:QFR} we present the main result of the paper, that of a general class of quantum fluctuation relations for dynamical observables. Importantly, these fluctuation relations hold true also for the Doob dynamics, despite a lack of manifest symmetries at the trajectory level. Three explicit examples are discussed in Section \ref{sec:ex} 
in order to illustrate properties of our general findings. The key points are summarised in the concluding Section \ref{sec:conc}.

\section{Thermodynamics of quantum jump trajectories}
\label{sec:traj}

Consider a quantum system described by a finite-dimensional Hilbert space experiencing a dissipative dynamics due to the interaction with an environment and such that memory effects in the time-evolution are negligible.
The dynamics of such a Markovian open quantum system is generally described by a master equation  $\partial_t \varrho_t = \mathcal{L}(\varrho_t)$, whose solution $\varrho_t$ represents the density matrix of the system at any time $t$, and where the generator $\mathcal{L}$ is in diagonal Lindblad form \cite{breuer02a,rivas12a}
\begin{equation}\label{lind}
\mathcal{L} (\varrho)= -i [H, \varrho ] + \sum_{\mu} \left( L_\mu \varrho L_\mu^\dag - \frac{1}{2}\{ L_\mu^\dag L_\mu , \varrho \} \right).
\end{equation}
The first part of the generator, $\mathcal{H}(\cdot)=-i [H, (\cdot) ]$, corresponds to the unitary evolution with a certain Hamiltonian operator $H$, while $\mathcal{D}(\cdot)= \sum_{\mu}  \left( L_\mu \cdot L_\mu^\dag - \frac{1}{2}\{ L_\mu^\dag L_\mu , \cdot \} \right)$ corresponds to the dissipative term. The operators $L_\mu$ are called Lindblad operators and describe the action of the environment on the system.
 
Without any access to the environment, this is the most accurate description of the dynamics of the system. Let us consider instead the case where we have a partial experimental access to the environment. In particular, let us consider an unraveling of the master equation in terms of quantum jump trajectories \cite{plenio98a,gardiner04a}, assuming we are able to detect all the jumps (and also to distinguish the different kind of jumps) by a continuous monitoring of the environment through a set of detectors.

More specifically, a quantum jump trajectory $\omega_t$ is completely specified by the sequence of jumps occurred, with jump $j$ labelled by the kind of jump $\mu_j$, and the jump time $t_j$,
\begin{equation}
\omega_t = (\mu_1,t_1,\mu_2,t_2, \ldots, \mu_n,t_n;t).
\end{equation}
Each variable $\mu_j$ is an integer taking values between $1$ and $N_L$, corresponding to different jump operators in the Lindbladian.
 Formally, if the system is initially in the pure state $\varrho_0 = \ketbra{\psi_0}{\psi_0}$, one can write 
\begin{equation}
\varrho_t = \sum_{n=0}^\infty \sum_{\mu_1=1}^{N_L}\ldots \sum_{\mu_n=1}^{N_L} \int_0^t  \mathrm{d}t_n   \ldots \int_0^{t_2}\!  \mathrm{d}t_1     \ketbra{\psi_t(\omega_t)}{\psi_t(\omega_t)},
\end{equation}
where $\ket{\psi_t(\omega_t)}$ is unraveled as follows
\begin{equation}
\ket{\psi_t(\omega_t)} = \mathrm{e}^{-i H_{\rm{eff}} (t-t_n)} L_{\mu_n} \ldots  L_{\mu_1} \mathrm{e}^{-i H_{\rm{eff}} t_1} \ket{\psi_0},
\end{equation}
in terms of the jump operators and of the effective Hamiltonian $H_{\rm{eff}}$ 
\begin{equation}
H_{\rm{eff}} = H - \frac{i}{2}\sum_{\mu=1}^{N_L} L_\mu^\dagger L_\mu .
\end{equation}
Note that the unravelling above is one of the possible Dyson expansions of the exponential of the Lindbladian, one chosen in terms of events given by the action of the jump operators. 

Each $\ket{\psi_t(\omega_t)}$ above represents the unnormalised state of the system conditioned on the sequence of jumps corresponding to trajectory $\omega_t$. One can compute the probability density of a certain specific sequence as
\begin{equation}
P(\omega_t) = \braket{\psi_t(\omega_t)}{\psi_t(\omega_t)}.
\end{equation}
In this framework, our aim is to discuss the statistics of a generic dynamical observable,
\begin{equation}
\underline{K}(\omega_t)=\sum_{\mu}Q_{\mu}(\omega_t)\valpha_{\mu} , 
\end{equation}
 whose components are linear combinations of the different number of jumps of $\mu$ type, $Q_{\mu}$, with vector coefficients $\valpha_{\mu}$. Such statistics is completely described by  the probability distribution $P(\underline{K})$ that is obtained summing $P(\omega_t)$ over all trajectories where $\underline{K}$ has a specific value, namely $P(\underline{K})= \sum_{\omega_t}P(\omega_t)\delta(\underline{K}(\omega_t)-\underline{K})$. The same information can be retrieved by the moment generating function $Z_{\underline{\lambda}}= \sum_{\{\underline{K} \}} \mathrm{e}^{-\vlambda^T \cdot \vK } P(\underline{K})$ that is conveniently represented as $Z_{\underline{\lambda}}= \mathrm{Tr}[\mathrm{e}^{t\mathcal{L}_{\underline{\lambda}}} (\varrho_0)]$ in terms of a tilted generator \cite{esposito09a,garrahan10a}
\begin{equation}
\mathcal{L}_{\vlambda}(\varrho)= \mathcal{H}(\varrho) + \mathcal{D_{\vlambda}}(\varrho) ,
\end{equation}
with 
\begin{equation}
\mathcal{D}_{\vlambda}(\varrho)= \sum_{\mu=1}^{N_L} \left( \mathrm{e}^{-\vlambda^T\cdot\valpha_\mu} L_\mu \varrho L_\mu^\dag - \frac{1}{2}\{ L_\mu^\dag L_\mu , \varrho \} \right).
\end{equation}
Here and in the following we indicate row vectors as $\underline{v}^{T}$ and the dot $\cdot$ is the usual product of matrices.
Under fairly general assumptions (see for instance \cite{jaksic14a,garrahan10a}), the statistics at long times is dominated by the largest eigenvalue of the tilted generator $\theta(\underline{\lambda})$,
\begin{equation}
\theta(\underline{\lambda}) = \lim_{t \to \infty} \frac{1}{t} \log Z_{\underline{\lambda}},
\end{equation}
the so-called scaled cumulant generating function (SCGF) \cite{touchette09a}. The name indicates that by taking derivatives of any order in $\lambda$ one can evaluate all the cumulants of the observable's probability distribution. The corresponding right and left eigenmatrices of $\mathcal{L}_{\vlambda}$,  $\mathcal{L}_{\vs}[r_{\vs}]=\theta(\vs)r_{\vs}$ and $\mathcal{L}_{\vs}^*[\ell_{\vs}]=\theta(\vs)\ell_{\vs}$, are denoted as $r_{\underline{\lambda}}$ and $\ell_{\underline{\lambda}}$, respectively, and are normalized as follows $\mathrm{Tr}[\ell_{\underline{\lambda}} \cdot r_{\underline{\lambda}}] = \mathrm{Tr}[ r_{\underline{\lambda}}]=1$. 

As one can easily check, the dynamics described by the tilted generator is not physical, in the sense that it does not preserve the trace. However, it is possible to find a proper physical dynamics generating the same biased statistics of the chosen observable for long times \cite{carollo18b}. This is the open quantum version of the so-called Doob dynamics \cite{jack10a,chetrite15b}, whose generator is defined as follows in terms of the tilted one \cite{carollo18b} 
\begin{equation}\label{doob}
\mathcal{L}_{\underline{s}}^{\rm Doob} (\cdot)= W_{\underline{s}} \circ \mathcal{L}_{\underline{s}} \circ W_{\underline{s}}^{-1} (\cdot) - \theta(\underline{s}) (\cdot),
\end{equation}
where $W_{\underline{s}} (\cdot)= \ell_{\underline{s}}^{1/2} (\cdot) \ell_{\underline{s}}^{1/2}$ . Here and in the following, we use the label $\vlambda$ to indicate a general biasing, while we use the variable $\underline{s}$ to indicate a physical field. In this respect, the Doob dynamics can be interpreted as the proper physical dynamics of the subset of trajectories leading to a given fluctuation of the chosen observable. In particular, while the rare fluctuations at some non-zero $\vlambda$ are rare in the original dynamics, they become typical in the Doob dynamics. In order to study the statistics of the relevant observable one can repeat the same procedure and tilt again the generator \eqref{doob}, obtaining
\begin{equation}
\mathcal{L}_{\vlambda,\vs}^{\rm Doob}[\cdot]=W_{\underline{s}} \circ  \mathcal{L}_{\vlambda + \vs} \circ W^{-1}_{\underline{s}} (\cdot)-\theta(\vs)(\cdot)\, .
\end{equation}
The spectrum of this tilted operator encodes the fluctuations of the trajectory observable $\underline{K}$ when the underlying dynamics is the Doob rather than the original one.

\section{Quantum Fluctuation relations due to symmetries}
\label{sec:QFR}

In this Section, following the rationale of our previous work dealing with classical stochastic processes \cite{marcantoni20a}, we derive a class of fluctuation relations by looking at the properties of $\theta(\underline{\lambda})$. A further ingredient needed for such purpose is a transformation $\mathcal{R}$ that is bijective in the space of quantum trajectories. 

For the sake of simplicity, we focus on those transformations that act locally in time, namely we consider a time-independent permutation of the jump types $\mu \to R\mu$. This in turn induces a transformation at the trajectory level
\begin{align*}
\omega_t &= (\mu_1,t_1,\mu_2,t_2, \ldots, \mu_n,t_n;t)\\ 
&\quad\quad\quad\quad\quad\quad   \mathcal{R}\big\downarrow \\
\mathcal{R}\omega_t &= (R\mu_1,t_1,R\mu_2,t_2, \ldots, R\mu_n,t_n;t)
\end{align*}
provided the new sequence of jumps is compatible with the dynamics. The transformation can be represented at the level of the quantum generator as a permutation of the jump operators, namely a linear transformation $\mathcal{V}$ such that $\mathcal{V}(L_{\mu})= L_{R\mu}$. In particular, given $\mathcal{V}$, the probability of the modified trajectory $\mathcal{R}\omega_t$ reads
\begin{align}\label{prob}
&P(\mathcal{R}\omega_t)= \bra{\psi_0} T^{\dag}_{t_1} L^\dag_{R\mu_1} \ldots T^{\dag}_{t-t_n}T_{t-t_n}\ldots L_{R\mu_1}T_{t_1}\ket{\psi_0} = \nonumber \\
&=\bra{\psi_0} T^{\dag}_{t_1} \mathcal{V}( L^\dag_{\mu_1}) \ldots T^{\dag}_{t-t_n}T_{t-t_n}\ldots \mathcal{V} (L_{\mu_1}) T_{t_1}\ket{\psi_0}
\end{align}
where we have defined $T_a=\mathrm{e}^{-i H_{\rm{eff}}a} $ and assumed $\mathcal{V}$ to be hermiticity preserving.
In order to find the fluctuation relation, we require the initial dynamics to have a symmetry, namely $P_0(\omega_t)=P_0(\mathcal{R}\omega_t)$. This holds true if the map $\mathcal{V}$ is unitary, so that it admits a representation $\mathcal{V}(\cdot)= V^\dag \cdot V$ with $V^\dag = V^{-1}$, and leaves the Hamiltonian of the system invariant $\mathcal{V}(H)=H$. More precisely, one has also to consider a symmetry of the initial state (which nevertheless is irrelevant in the long-time limit we discuss in the following) $V \ket{\psi_0}=\mathrm{e}^{i\phi}\ket{\psi_0}$.  Indeed, from \eqref{prob} one has
\begin{align*}
&=\bra{\psi_0} T^{\dag}_{t_1} \mathcal{V}( L^\dag_{\mu_1}) \ldots T^{\dag}_{t-t_n}T_{t-t_n}\ldots \mathcal{V} (L_{\mu_1}) T_{t_1}\ket{\psi_0} = \\
&=\bra{\psi_0} V^{\dag}  \widetilde{T}^{\dag}_{t_1} L^\dag_{\mu_1}  \ldots \widetilde{T}^{\dag}_{t-t_n}\widetilde{T}_{t-t_n}\ldots L_{\mu_1} \widetilde{T}_{t_1} V\ket{\psi_0} = \\
&= \bra{\psi_0} T^{\dag}_{t_1} L^\dag_{\mu_1} \ldots T^{\dag}_{t-t_n}T_{t-t_n}\ldots L_{\mu_1}T_{t_1}\ket{\psi_0}  = P(\omega_t),
\end{align*}
where we defined $\widetilde{T}_{a}= V T_a V^\dag$ and used the invariance of the Hamiltonian to say that $T_a=\widetilde{T}_{a}$. Note that the unitary transformation $\mathcal{V}$ preserves the Hilbert-Schmidt norm of the Lindblad operators, $\mathrm{Tr}[L^\dag_{\mu} L_\mu]= \mathrm{Tr}[L^\dag_{R\mu} L_{R\mu}] $. This fact can be interpreted as a symmetry on the jump rates, resembling the condition discussed in the classical case \cite{marcantoni20a}.

With all the previously discussed machinery, by choosing an observable $\underline{K}(\omega_t)$ that transforms under the permutation of jumps as $\underline{K}(\mathcal{R} \omega_t)=U\cdot \underline{K}(\omega_t)$, one can derive the following symmetry of the tilted generator
\begin{equation}\label{symm}
\mathcal{L}_{\underline{\lambda}} = \mathcal{V} \circ \mathcal{L}_{(U^{-1})^T \cdot \underline{\lambda}} \circ \mathcal{V}^{-1}.
\end{equation}
The relation \eqref{symm} on the dissipative part can be easily verified as follows
\begin{align}
\mathcal{D}_{\underline{\lambda}}=&\sum_{\mu}  \left( \mathrm{e}^{-\vlambda^T\cdot\valpha_\mu} L_\mu \varrho L_\mu^\dag - \frac{1}{2}\{ L_\mu^\dag L_\mu , \varrho \} \right) = \nonumber \\
=& \sum_{\mu} \left( \mathrm{e}^{-\vlambda^T\cdot\valpha_\mu}V L_{R\mu}  V^{-1} \varrho V L_{R\mu}^\dag V^{-1} \right. \nonumber \\
& \quad - \left. \frac{1}{2}V \{ L_{R\mu}^\dag L_{R\mu} , V^{-1} \varrho V \} V^{-1} \right) = \nonumber \\
=& \sum_{\mu} \left( \mathrm{e}^{-\vlambda^T\cdot\valpha_{R^{-1}\mu}}V L_{\mu}  V^{-1} \varrho V L_{\mu}^\dag V^{-1} \right.  \nonumber \\
& \quad - \left. \frac{1}{2}V\{ L_{\mu}^\dag L_{\mu} , V^{-1} \varrho V \} V^{-1} \right) = \nonumber \\
=& \sum_{\mu}  \left( \mathrm{e}^{-[(U^{-1})^T\cdot \vlambda]^T \cdot \valpha_\mu}V L_{\mu}  V^{-1} \varrho V L_{\mu}^\dag V^{-1} \right.  \nonumber \\
& \quad -\left. \frac{1}{2}V^{-1}\{ L_{\mu}^\dag L_{\mu} , V^{-1} \varrho V \} V \right)\nonumber \\
=& \mathcal{V} \circ \mathcal{D}_{ (U^{-1})^T  \cdot \underline{\lambda}} \circ \mathcal{V}^{-1}
\end{align}
where in the second-to-last line we used the fact that $\alpha_{R^{-1}\mu}=U^{-1}\cdot \alpha_\mu$, as can be seen from the definition of $\vK(\omega_t)$, the relation $\underline{K}(\mathcal{R} \omega_t)=U\cdot \underline{K}(\omega_t)$ and the identity $Q_\mu(\omega_t)=Q_{R\mu}(R\omega_t)$ (see also Ref.\cite{marcantoni20a}). Finally, equation \eqref{symm} follows from the assumed invariance of the Hamiltonian. 
The assumption of a symmetric Hamiltonian can be relaxed if the Hamiltonian part of the generator commutes with the dissipative tilted one, $\mathcal{H}=\mathcal{D_{\vlambda}}^{-1}\circ \mathcal{H} \circ \mathcal{D_{\vlambda}}$, so that one can diagonalize them separately and consider the symmetry only on the dissipative part. If this is the case, the dynamics becomes just classical hopping between the eigenstates of the Hamiltonian.
Notice that from \eqref{symm} one can infer a symmetry on the long-time fluctuations of the observable $\underline{K}$, as described by the SCGF $\theta(\lambda)=\theta(-\lambda)$.

More interestingly, from the previous result one can go further and find a fluctuation relation in a dynamics where there is no initial symmetry on the rates. This is the dynamics obtained by applying the Doob transform to the original one. The Doob dynamics for a given value of the biasing field, say $\vs$, breaks explicitly the symmetry; but nevertheless there is still trace of it in the statistics of fluctuations. 
For long times, the SCGF, $\theta(\vs)$, is given by the largest eigenvalue of $\mathcal{L}_{\vs}$, so that $\mathcal{L}_{\vs}^*[\ell_{\vs}]=\theta(\vs)\ell_{\vs}$ and $\mathcal{L}_{\vs}[r_{\vs}]=\theta(\vs)r_{\vs}$, where $\ell_{\vs}$ and $r_{\vs}$ are the corresponding left and right eigenmatrices, normalized such that ${\rm Tr}[\ell_{\vs}\cdot r_{\vs}]={\rm Tr}[r_{\vs}]=1$. By tilting the Doob dynamics we get
\beq
\mathcal{L}_{\vlambda,\vs}^{\rm Doob}[\cdot]=\ell_s^{1/2} \mathcal{L}_{\vlambda + \vs}[\ell_{\vs}^{-1/2}(\cdot)\ell_{\vs}^{-1/2}] \ell_{\vs}^{1/2}-\theta(\vs)(\cdot)\, ,
\label{tiltD}
\eeq
such that the Doob transform is given for $\vlambda=0$, which corresponds to a proper (probability preserving) dynamics $\mathcal{L}_{\vlambda=0,\vs}^{\rm Doob,*}[\mathbbm{1}]=0$. Then, by using the transformation $W_{\vs}(\cdot)=\ell_{\vs}^{1/2}(\cdot)\ell_{\vs}^{1/2}$, we get from Eqs. \eqref{symm} and \eqref{tiltD} the following similarity relation
\beq \label{simDoob}
\mathcal{L}_{\vlambda,\vs}^{\rm Doob}[\cdot]=A_{\vs}\circ \mathcal{L}_{(U^{-1})^T \cdot(\vlambda+\vs)-\vs, \vs}^{\rm Doob}[\cdot] \circ A_{\vs}^{-1},
\eeq
with $A_{\vs}=W_{\vs}\circ \mathcal{V} \circ W_{\vs}^{-1}$. This fact can be easily proved in a few steps
\begin{align*}
&\mathcal{L}_{\vlambda,\vs}^{\rm Doob} = W_{\vs}\circ \mathcal{L}_{\vlambda + \vs} \circ W_{\vs}^{-1} - \theta(\vs) \\
&\quad \overset{(12)}{=} W_{\vs}\circ \mathcal{V} \circ \mathcal{L}_{(U^{-1})^T \cdot (\vlambda + \vs)} \circ \mathcal{V}^{-1} \circ W^{-1}_{\vs} - \theta(\vs) \\
&\quad \overset{(14)}{=} W_{\vs}\circ \mathcal{V} \circ W^{-1}_{\vs} \circ \mathcal{L}_{(U^{-1})^T \cdot(\vlambda+\vs)-\vs, \vs}^{\rm Doob} \circ W^{-1}_{\vs}\circ \mathcal{V} \circ W_{\vs}.
\end{align*}
It is important to notice that the Doob dynamics can be recast in the Lindblad form \cite{carollo18b} with effective Hamiltonian $H_{\vs}$ and jump operators $L_\mu^{\vs}$
\begin{equation}\label{newop}
H_{\vs}= \frac{1}{2} \ell_{\vs}^{1/2} H_{\rm{eff}} \ell_{\vs}^{-1/2} + h.c. ,\quad L_\mu^{\vs} =  \mathrm{e}^{-\frac{1}{2}\vs^T\cdot\valpha_\mu} \ell_{\vs}^{1/2} L_\mu \ell_{\vs}^{-1/2}.
\end{equation}
The tranformation representing the permutation of jumps in this setting cannot be unitary, in general, because $$\mathrm{Tr}[(L_\mu^{\vs})^{ \dag} L_\mu^{\vs}] = \mathrm{e}^{-\vs^T\cdot\valpha_\mu} \mathrm{Tr}[L_\mu^{\dag} \ell_{\vs} L_\mu  \ell^{-1}_{\vs} ] \neq \mathrm{Tr}[(L_{R\mu}^{\vs})^{ \dag} L_{R\mu}^{\vs}] , $$
so that there is no initial symmetry in the Doob dynamics, i.e. $P_{\vs}(\mathcal{R}\omega_t)\neq P_{\vs} (\omega_t)$.
However, identifying $\vs$ with a constant physical field, the relation \eqref{simDoob} above implies the fluctuation relation 
\beq \label{FR}
\theta_{\vs}(\vlambda)=\theta_{\vs}[ (U^{-1})^T\cdot(\vlambda+\vs)-\vs]\, ,
\eeq
which is the main result of this paper.

\section{Examples}
\label{sec:ex}

In the following we discuss specific examples of the general fluctuation relation obtained above in three systems of increasing complexity. The first example we consider is related to the depolarising dynamics of a single qubit. The second example is that of a couple of qubits, while the third one corresponds to a simple, yet many-body, problem.

\subsection{Single qubit}
Consider the following tilted generator
\begin{align}
\mathcal{L}_\lambda(\varrho) =& -i [\Omega \sigma_x, \varrho ] + \gamma_- \left(\mathrm{e}^{\lambda} \sigma_- \varrho \sigma_+ - \frac{1}{2}\{ \sigma_+ \sigma_- , \varrho \} \right)+ \nonumber \\
&+ \gamma_+ \left(\mathrm{e}^{-\lambda} \sigma_+ \varrho \sigma_- - \frac{1}{2}\{ \sigma_- \sigma_+ , \varrho \} \right),
\end{align}
where $\sigma_x$ is the usual Pauli matrix, $\sigma_+= \ketbra{2}{1}$, $\sigma_- = \ketbra{1}{2}$, with $\ket{1}$ and $\ket{2}$ eigenstates of $\sigma_z$ corresponding to the eigenvalues $-1$ and $1$, respectively, and $\gamma_+, \gamma_-$ are two positive damping rates. For $\lambda=0$ this is a Lindblad generator for the dynamics of a qubit, while for $\lambda \neq 0$ it describes the statistics of the observable $K= K_+ - K_-$, that is the difference between the number $K_+$ of jumps $\ket{1} \to \ket{2}$ and the number of the jumps $K_-$ in the opposite direction. The only nontrivial transformation $R$ is the exchange of the two labels $1 \leftrightarrow 2$ that is represented in the Hilbert space by the action of the operator $V=\sigma_x$.  The observable $K$ changes sign under the permutation $R$, so that $U=-1$ in this case. The Hamiltonian is invariant and the symmetry \eqref{symm} holds true if the rates are equal $\gamma_+=\gamma_-=\gamma$. In particular one has for the scaled cumulant generating function the symmetry $\theta(\lambda)=\theta(-\lambda)$. This can be checked by computing the eigenvalues of $\mathcal{L}_{\lambda}$. For the particular case of $4\Omega^2=\gamma^2$, one explicitly has for the scaled cumulant generating function $\theta(\lambda)= \gamma \Big(\cosh^{1/3}(\lambda)-1\Big)$. The corresponding left and right eigenmatrices read
\begin{align}
\ell_\lambda=& \frac{\sinh(\lambda)}{3  \cosh^{2/3}(\lambda) \Big( \cosh^{2/3}(\lambda)-1 \Big)} \times \nonumber\\
&\times \begin{pmatrix}
\mathrm{e}^{\lambda} -  \cosh^{1/3}(\lambda)    & i \Big(1- \cosh^{2/3}(\lambda) \Big) \\
 -i \Big(1- \cosh^{2/3}(\lambda) \Big)                   &  \cosh^{1/3}(\lambda)-\mathrm{e}^{-\lambda}
\end{pmatrix},
\end{align}
\begin{align}
r_\lambda=& \frac{1}{2 \sinh(\lambda)} \times \nonumber \\
&\times \begin{pmatrix}
\cosh^{1/3}(\lambda)-\mathrm{e}^{-\lambda}       & i \Big(1- \cosh^{2/3}(\lambda) \Big) \\
 -i \Big(1- \cosh^{2/3}(\lambda) \Big)                   & \mathrm{e}^{\lambda} -  \cosh^{1/3}(\lambda)
\end{pmatrix}.
\end{align}
It is convenient to parametrize the matrix $\ell_{\lambda}^{1/2}$ as follows
\begin{equation}
\ell_{\lambda}^{1/2}= 
\begin{pmatrix}
\alpha &-i\delta \\
i\delta & \beta
\end{pmatrix},
\end{equation}
so that its inverse $\ell_{\lambda}^{-1/2}$ reads
\begin{equation}
\ell_{\lambda}^{-1/2}= \frac{1}{\alpha\beta -\delta^2}
\begin{pmatrix}
\beta & i\delta \\
-i\delta & \alpha
\end{pmatrix}.
\end{equation}
The expression of $\alpha,\beta,\delta$ is somewhat involved and given explicitly in Appendix \ref{app: explicit}.

The generator of the Doob transformed dynamics reads
\begin{align}\label{Doob}
\mathcal{L}_s^{\rm{Doob}}(\varrho) &= -i\big[ H_s, \varrho \big] + \Big(  L_1^s \varrho (L_1^s)^\dag - \frac{1}{2} \big\{ (L_1^s)^\dag L_1^s, \varrho \big\} \Big) \nonumber \\
&\quad + \Big(  L_2^s \varrho (L_2^s)^\dag - \frac{1}{2} \big\{ (L_2^s)^\dag L_2^s, \varrho \big\} \Big) .
\end{align}
where the Hamiltonian and jump operators are given by, cf. Eq. \eqref{newop},
\begin{align}
H_s &= \frac{\gamma}{4}\frac{\alpha^2 + \beta^2 + 2\delta^2}{\alpha\beta - \delta^2} \sigma_x =\\
&=\frac{\gamma}{2} \frac{\big|\sinh(s)\big|}{\sqrt{\Big( \cosh^{2/3}(s)-1 \Big)\Big( \cosh^{2/3}(s)+2\Big)}} \sigma_x  ,
\end{align}
\begin{equation}
\label{L1s1q}
L_1^s = \sqrt{\gamma} \mathrm{e}^{s/2} \ell_{s}^{1/2} \sigma_- \ell_{s}^{-1/2} = \frac{\sqrt{\gamma} \mathrm{e}^{s/2} }{\alpha\beta - \delta^2} 
\begin{pmatrix}
-i\beta \delta & \delta^2 \\
\beta^2 & i\beta\delta
\end{pmatrix}
\end{equation}
\begin{equation}
\label{L2s1q}
L_2^s = \sqrt{\gamma} \mathrm{e}^{-s/2} \ell_{s}^{1/2} \sigma_+ \ell_{s}^{-1/2} = \frac{\sqrt{\gamma} \mathrm{e}^{-s/2} }{\alpha\beta - \delta^2} 
\begin{pmatrix}
-i\alpha \delta & \alpha^2 \\
\delta^2 & i\alpha\delta
\end{pmatrix}.
\end{equation}
Note that the two jump operators describe different processes with respect to the initial ones $\sigma_-$ and $\sigma_+$, therefore, in order to interpret the fluctuation relation one has to conceive an experiment where the number of jumps of the first type, given by \eqref{L1s1q}, can be counted and distinguished from the number of jumps of the second type, given by \eqref{L2s1q}. 

One can tilt also the Doob generator and find the scaled cumulant generating function $\theta_{s}(\lambda)$
\begin{equation}
\theta_{s}(\lambda) = \theta(\lambda +s) - \theta(s)= \gamma \cosh^{1/3}(\lambda+s) - \gamma \cosh^{1/3}(s).
\label{tsl}
\end{equation}
Therefore, the Doob dynamics satisfies the fluctuation relation \eqref{FR}, where $s$ has the role of the physical field. Recalling that $U=-1$ in this case one has 
\begin{equation}
\theta_s(\lambda)= \theta_s(-\lambda-2s),
\end{equation}
which can be easily checked to be correct from (\ref{tsl}) by inspection.

Before concluding this example we make a further observation. One could also consider the dynamics with an Hamiltonian $H=\sigma_z$ instead of $\sigma_x$. In such a case, one can easily verify that the unitary part commutes with the dissipative one and can be diagonalized separately. In this case the dynamics reduces to a classical hopping between the two states $\ket{1}$ and $\ket{2}$, and the chosen observable is not extensive in time in the proposed dynamics.

\subsection{Two qubits}
Another simple example is a two-spin system with hopping Hamiltonian $H= g(\sigma^+_A \sigma^-_B + \sigma^+_B \sigma^-_A)$ and local dissipators. Here, the operator $\sigma^{\pm}_{A},\sigma^{\pm}_{B} $ refer to the ladder operators pertaining to the spin $A$ or $B$, respectively, and $g$ is just a coefficient describing the strength of the interaction. By labelling the four possible states of the local basis as follows, $\ket{\downarrow \downarrow}= \ket{1}, \ket{\downarrow \uparrow}= \ket{2}, \ket{\uparrow \downarrow}= \ket{3}, \ket{\uparrow \uparrow}= \ket{4}$, one can consider the eight jump operators 
\begin{align*}
& L_1= \sqrt{\alpha} \ketbra{2}{1}, \quad L_5=\sqrt{\alpha} \ketbra{3}{1} \\
& L_2=\sqrt{\alpha} \ketbra{1}{2}, \quad L_6=\sqrt{\alpha} \ketbra{1}{3} \\
& L_3=\sqrt{\alpha} \ketbra{2}{4}, \quad L_7=\sqrt{\alpha} \ketbra{3}{4} \\
& L_4=\sqrt{\alpha} \ketbra{4}{2}, \quad L_8=\sqrt{\alpha} \ketbra{4}{3}
\end{align*} 
that describe single spin flip with equal rates for the transitions $\downarrow\,\to\,\uparrow$ and $\uparrow\,\to\,\downarrow$. In this setting one can consider for instance the transformation $R$ that exchanges $2 \leftrightarrow 3$, so that $L_{n}, n\in{1,2,3,4}$ is mapped into $L_{n+4}$ and viceversa, and consider as an observable the difference between jumps dealing with the state $\ket{2}$ and jumps involving $\ket{3}$. Also in this case $U=-1$ and the Hamiltonian turns out to be invariant, because it can be equivalently rewritten as $H= g (\ketbra{3}{2} + \ketbra{2}{3})$. Therefore the fluctuation relation \eqref{FR} holds true. 
To better see this, let us compute explicitly the scaled cumulant generating function from the tilted generator
\begin{align}
\mathcal{L}_\lambda (\varrho) &= -i[H, \varrho] + \sum_{k=1}^4 \Big( \mathrm{e}^{-\lambda} L_k \varrho L_k^\dag -\frac{1}{2} \Big\{ L_k^\dag L_k, \varrho \Big\} \Big) \nonumber \\
&\quad +\sum_{k=5}^8 \Big( \mathrm{e}^{\lambda} L_k \varrho L_k^\dag -\frac{1}{2} \Big\{ L_k^\dag L_k, \varrho \Big\} \Big).
\end{align}
One can notice that for $\lambda=0$ the dynamics is unital and has a unique stationary state, which is the totally mixed one $\mathbbm{1}_4$. For $\lambda\neq 0$ the identity is no longer preserved, however, it turns out that the six-dimensional space spanned by the matrices $\{\ketbra{1}{1}, \ketbra{2}{2},\ketbra{3}{3},\ketbra{4}{4}, \ketbra{2}{3}, \ketbra{3}{2}\}$ is invariant under the action of the generator. Therefore, we start to look for the highest eigenvalue in this subspace. After some algebra, one finds the six eigenvalues and, in particular, the highest one is
\begin{widetext}
\begin{align}
\theta(\lambda) = -2\alpha + \sqrt{2\alpha^2 \cosh(2\lambda)- 2g^2 +2\sqrt{\alpha^4\cosh^2(2\lambda) + g^4 + 2\alpha^2 g^2}}.
\end{align}
\end{widetext}
By completing the diagonalization in the complementary subspace one can check that this is indeed the overall highest eigenvalue. As expected, it obeys the relation $\theta(\lambda)=\theta(-\lambda)$. In order to construct the Doob transform we need the right and the left eigenmatrices, that have the following simple structure 
\begin{equation}
r_\lambda =
\begin{pmatrix}
a &0 &0 &0 \\
0 &c &im &0 \\
0 &-im & d &0 \\
0 &0 &0 &b
\end{pmatrix} , \quad 
\ell_\lambda = \eta
\begin{pmatrix}
a &0 &0 &0 \\
0 &c &-im &0 \\
0 &im & d &0 \\
0 &0 &0 &b
\end{pmatrix},
\end{equation}
where the real parameters $a,b,c,d,m,\eta$ read as follows (here $\gamma_\lambda=\theta(\lambda)+2\alpha$)
\begin{align*}
&a=b= \frac{\gamma_\lambda}{2(\gamma_\lambda + 2\alpha\cosh(\lambda))}, \\
&c= \frac{2\alpha^2 (\mathrm{e}^{2\lambda} + 1) - \gamma_\lambda^2}{4\alpha\sinh(\lambda)((\gamma_\lambda + 2\alpha\cosh(\lambda)))} ,\\
&d= \frac{-2\alpha^2 (\mathrm{e}^{-2\lambda} + 1) + \gamma_\lambda^2}{4\alpha\sinh(\lambda)((\gamma_\lambda + 2\alpha\cosh(\lambda)))}, \\
&m= \frac{2\alpha g \cosh(\lambda) -g \gamma_\lambda }{2\alpha\gamma_\lambda \sinh(\lambda)} ,\\
&\eta = \frac{1}{2a^2 + c^2 + d^2 - 2m^2}.
\end{align*}
In computing the coefficients, we have already assumed $\mathrm{Tr}[r_\lambda]=a+b+c+d=1$, while the further condition $\mathrm{Tr}[\ell_\lambda\cdot r_\lambda]=1$ is ensured by the normalization $\eta$.
The particular structure can be easily understood. First of all, they are both block-diagonal because they belong to the six-dimensional subspace previously mentioned. Moreover, the element ${11}$ is equal to the element ${44}$ because there is a symmetry in the generator under the exchange $\ket{4} \leftrightarrow \ket{1}$. Finally one can argue that $\ell_\lambda$ has the same elements of $r_\lambda$, up to a normalization $\eta$ and upon a change $g \to -g$, because the dual generator is equivalent to the original one with $g$ replaced by $-g$.
The matrix $\ell_\lambda^{1/2}$ and its inverse, both used for the Doob, have also a similar structure
\begin{align}
&\ell_\lambda^{1/2} = \sqrt{\eta}
\begin{pmatrix}
\sqrt{a} &0 &0 &0 \\
0 &A &iC &0 \\
0 &-iC & B &0 \\
0 &0 &0 &\sqrt{a}
\end{pmatrix} , \\
&\ell_\lambda^{-1/2} = \frac{1}{\sqrt{\eta}}
\begin{pmatrix}
\frac{1}{\sqrt{a}} &0 &0 &0 \\
0 &\frac{B}{AB-C^2} &\frac{-iC}{AB-C^2} &0 \\
0 &\frac{iC}{AB-C^2} & \frac{A}{AB-C^2} &0 \\
0 &0 &0 &\frac{1}{\sqrt{a}}
\end{pmatrix},
\end{align}
where the explicit expression of the parameters $A,B,C$ is presented in Appendix \ref{app: ex2}. 
Therefore, it turns out that also in this case the Doob Hamiltonian is algebraically equivalent (but with a different coefficient; see equation \eqref{newop}) 
\begin{equation}
H_s = \frac{g}{2}\frac{A^2 + B^2 + 2C^2}{AB-C^2} \big( \ketbra{2}{3} + \ketbra{3}{2} \big)
\end{equation}
while the jump operators are rotated. Moreover, they have different rates and there is no unitary transformation permuting the jump operators. For instance,
\begin{align*}
& L_1^s= \mathrm{e}^{-s/2} \frac{\sqrt{\alpha}}{\sqrt{a}} \Big( A\ketbra{2}{1} -iC \ketbra{3}{1} \Big), \\
& L_5^s= \mathrm{e}^{s/2}\frac{\sqrt{\alpha}}{\sqrt{a}}  \Big( B\ketbra{3}{1} +iC \ketbra{2}{1} \Big),
\end{align*} 
and $\mathrm{Tr}[(L_1^s)^{\dag} L_1^s]\neq \mathrm{Tr}[(L_5^s)^{\dag} L_5^s]$. For completeness the other jump operators are reported in Appendix \ref{app: ex2}. The scaled cumulant generating function of the Doob dynamics, which satisfies the fluctuation relation \eqref{FR}, is then obtained by tilting with respect to the same combination of jumps.

\subsection{Spin chain}

Our final example is a simple many-body problem.
Consider a quantum spin chain, composed of $N$ spin-$1/2$, with periodic boundary conditions $\sigma^i_{N+1} = \sigma^i_{1}  $ and local jump operators at each site
\begin{equation}
H= -J \sum_{k=1}^N \sigma^z_k \sigma^z_{k+1}, \quad L^x_k = \sqrt{\gamma} \sigma^x_k, \quad L^y_k = \sqrt{\gamma} \sigma^y_k .
\end{equation}
We assume $N$ to be even, so that there is an equal number of even and odd sites.
We want to study the statistics of the difference between the number of jumps in even sites and the number of jumps in odd sites, namely the chosen dynamical observable is $K = K_{\rm{even}}- K_{\rm{odd}}$. The transformation $R$ we choose in this setting shifts the site $k$ into $k+1$, so that it switches even and odd sites and the $U=-1$ also in this case (one could equivalently choose $R$ as a shift by an arbitrary odd number of sites). The tilted generator corresponding to the observable therefore reads
\begin{align}
\mathcal{L}_\lambda (\varrho)&= -i[H, \varrho ] + \gamma \sum_{k=1}^{N/2} \Big( \mathrm{e}^{-\lambda} (\sigma^x_{2k} \varrho \, \sigma^x_{2k} +\sigma^y_{2k} \varrho \, \sigma^y_{2k}) -2\varrho  \Big) \nonumber \\
&\quad + \gamma \sum_{k=1}^{N/2} \Big( \mathrm{e}^{\lambda} (\sigma^x_{2k-1} \varrho\, \sigma^x_{2k-1} +\sigma^y_{2k-1} \varrho\, \sigma^y_{2k-1}) -2\varrho  \Big).
\end{align}
Following \cite{fossfeig17a}, a convenient parametrization of the density matrix is in terms of Pauli strings
\begin{equation}
\varrho = \sum_{\{m_1, \ldots m_N\}} \varrho_{m_1,\ldots,m_N} \,\sigma_1^{m_1} \otimes \sigma_2^{m_2} \otimes \ldots \otimes \sigma_N^{m_N},
\end{equation}
where each label $m_i$ can take four values $\{ 1, x,y,z \}$ and the matrix $\sigma^{1}$ is understood to be the identity $\mathbbm{1}$. The coefficients $\varrho_{m_1,\ldots,m_N} $ define a vector in a $4^N$ dimensional space so that the tilted generator $\mathcal{L}_\lambda$ inherits a matrix representation. Since the generator is hermiticity preserving, this matrix representation in the chosen basis has real entries. Moreover, one can notice that the dissipative part of the generator is already diagonal, in the sense that any element $\sigma_1^{m_1} \otimes \sigma_2^{m_2} \otimes \ldots \otimes \sigma_N^{m_N}$ is mapped into itself with some coefficient. Concerning the Hamiltonian part, the structure is a bit more complicated, however, one can notice that the overall number of operators $\sigma^x$ and $\sigma^y$ is conserved. This is because the Hamiltonian part of the generator transforms $\sigma^x$ into $\sigma^y$ (and viceversa) and $\sigma^z$ into $\mathbbm{1}$ (and viceversa). Therefore, the generator has a block diagonal structure where each block is labelled $d \in \{0,1,\ldots, N\}$ indicating the number of $x,y$ Pauli matrices in the list of indices $m_1,\ldots,m_N$. Each block $\mathcal{L}_\lambda^d$ acts on a Hilbert space of dimension $\frac{2^N N!}{d! (N-d)!}$ so that the largest one is for $d=N/2$.
Due to the block-diagonal structure, the spectrum of $\mathcal{L}_\lambda$ is obtained as the union of the eigenvalues in each bloch
\begin{equation}
sp(\mathcal{L}_\lambda)= \bigcup_{d=0}^N sp(\mathcal{L}^d_\lambda).
\end{equation}
The $2^N$ eigenvalues of $\mathcal{L}^0_\lambda$ can be easily obtained because the Hamiltonian part leaves each element in this subspace invariant (these are just products of identities and matrices $\sigma^z$) so that only the dissipative part contributes and the generator is therefore already diagonal. In particular, one immediately notices that the eigenmatrix $2^{-N} \mathbbm{1}_{2^N}$ corresponds to the eigenvalue $0$ for $\lambda=0$, while for generic $\lambda$ one has eigenvalue $2\gamma N [\cosh(\lambda)-1]$.
Since the set $\{ H, L^x_k, L^y_k \}$ generates the full algebra $\mathcal{M}_{2^N}(\mathbb{C})$ of the $2^N \times 2^N$ complex matrices, the Evans criterion \cite{evans77a} is satisfied and the generator for $\lambda=0$ has a unique steady state. All the other eigenvalues have strictly negative real part. Using the continuity of the spectrum with respect to $\lambda$ and the spectral gap for $\lambda=0$, one can heuristically argue that at least for $\lambda$ sufficiently small, the largest eigenvalue is still the one corresponding to the identity. This argument is however unsatisfactory because it does not allow us to predict if one can find level crossing at finite $\lambda$. Importantly, exploiting the particular structure of the generator, in Appendix \ref{app: ex3} we indeed show that the eigenvalue corresponding to the identity matrix is the one with the largest real part for any $\lambda$. Therefore, we have 
\begin{equation}
\theta(\lambda)= 2\gamma N [\cosh(\lambda)-1]\, ,
\end{equation}
with $r_\lambda = 2^{-N} \mathbbm{1}_{2^N} $ and $\ell_\lambda= \mathbbm{1}_{2^N} $. The Doob dynamics is a Lindblad dynamics with the same Hamiltonian $H$ and with the same jump operators up to a site-dependent factor, that discriminates between even and odd sites
\begin{equation}
L^s_{2k} = \mathrm{e}^{-s/2} L_{2k}, \quad L^s_{2k-1} = \mathrm{e}^{s/2} L_{2k-1}.
\end{equation}
Therefore, the field $s$ is responsible for the modification of the jump rates, so that the Doob dynamics is no longer symmetric under the translation by an odd number of sites. As a consequence, the scaled cumulant generating function obtained by tilting the Doob dynamics for the physical field $s$ obeys the fluctuation relation given by \eqref{FR}.

\section{Conclusion}
\label{sec:conc}

In summary, we presented a general recipe to build quantum fluctuation relations for dynamical observables. We used the formalism of quantum jump trajectories to describe the dynamics of Markovian open quantum systems and we studied the statistics of generic dynamical observables using techniques form large deviation theory. As a result, we obtained a generalisation to the quantum setting of a scheme previously discussed \cite{marcantoni20a} in the context of classical continuous-time Markov chains. As in the classical scheme, for open quantum systems we exploited the following facts: 
(i) Starting from a dynamics which has a certain symmetry at the trajectory level, we can define a second dynamics where this symmetry is broken by considering a tilt (or deformation) of the Lindbladian generator associated with some non-invariant observable; for the case of quantum jump unravellings, such observable correspond to some counting of quantum jumps. (ii) The tilted generator is associated with a dynamical ensemble where the probabilities of trajectories are exponentially tilted with respect to those original one; this tilted operator is however not a stochastic generator. (iii) Nevertheless, by means of a Doob transformation we can obtain a bona fide stochastic dynamics with the same ensemble of trajectories as in (ii); this means that in general we can always construct a pair of physically consistent dynamics where one is symmetric and the second one is non-symmetric and exponentially tilted with respect to the first. (iv) As long as the transformation on the trajectories induces a unique (trajectory-independent) transformation on the observable, the new dynamics displays a fluctuation relation in the statistics of this observable inherited from the symmetry properties of the original dynamics. 

Note the following: First, our construction is not based on time reversal, the usual symmetry relevant for standard fluctuation relations of current-like quantities. Second, for this open quantum case we do not rely on two-measurement schemes \cite{campisi11a}. Third, the relations we discussed are in principle observable in experimental setups that allow for a recording of the sequence of jump events, like in photon counting experiments. Fourth, for concreteness we focused on local in time transformations, but we envisage a generalisation of our scheme to symmetries of the trajectory ensemble that mix event times. This raises the intriguing possibility of a connection with the so-called ``retrodiction'' \cite{Tan2015} problem in quantum trajectories. Among this and other interesting connections, we aim to explore in the near future practical schemes to implement the ideas we described here in experimental setups.

\section*{Acknowledgments}
The research leading to these results has received funding from the European Union's Horizon 2020 research and innovation programme under the Marie Sklodowska-Curie Cofund Programme Athenea3I Grant Agreement No. 754446, from the European Regional Development Fund, Junta de Andaluc\'ia-Consejer\'ia de Econom\'ia y Conocimiento, Grant No. A-FQM-175-UGR18 and from the EPSRC Grant No. EP/R04421X/1. J.P.G. is grateful to All Souls College, Oxford, for support through a Visiting Fellowship during the time most of this work was carried out.

\appendix

\section{Explicit parameters used in the first example}
\label{app: explicit}

Here we present for completeness the expression of the three parameters $\alpha, \beta$ and $\delta$ defined in the first Example.

\begin{widetext}

\begin{align*}
\alpha &=  \sqrt{ \frac{\Big|\sinh(\lambda)\Big|}{6 \cosh^{2/3}(\lambda) \Big( \cosh^{4/3}(\lambda)-1 \Big) } } \times \\
& \quad \times \sqrt{\Big|\sinh(\lambda)\Big|\Big(2 \cosh^{2/3}(\lambda)+1\Big) + 2\, \mathrm{sign}(\lambda) \cosh^{1/3}(\lambda) \Big( \cosh^{4/3}(\lambda)-1 \Big) + \sqrt{\Big( \cosh^{2/3}(\lambda)-1 \Big)\Big( \cosh^{2/3}(\lambda)+2\Big)}} \\
\vspace{1cm} \\
\beta &= \sqrt{ \frac{\Big|\sinh(\lambda)\Big|}{6 \cosh^{2/3}(\lambda) \Big( \cosh^{4/3}(\lambda)-1 \Big) } } \times \\
& \quad \times \sqrt{\Big|\sinh(\lambda)\Big| \Big(2 \cosh^{2/3}(\lambda)+1\Big) - 2\, \mathrm{sign}(\lambda) \cosh^{1/3}(\lambda) \Big( \cosh^{4/3}(\lambda)-1 \Big) + \sqrt{\Big( \cosh^{2/3}(\lambda)-1 \Big)\Big( \cosh^{2/3}(\lambda)+2\Big)}} \\
\vspace{1cm} \\
\delta &= \mathrm{sign}(\lambda) \sqrt{ \frac{\Big|\sinh(\lambda)\Big|}{6 \cosh^{2/3}(\lambda) \Big( \cosh^{4/3}(\lambda)-1 \Big) }   \times \Bigg( \Big|\sinh(\lambda)\Big| - \sqrt{\Big( \cosh^{2/3}(\lambda)-1 \Big)\Big( \cosh^{2/3}(\lambda)+2\Big)} \Bigg) }
\end{align*}

\section{Details for the second example}
\label{app: ex2}

In this Appendix we present some computational details about the second example. In particular, the three parameters $A,B,C$ read
\begin{align*}
A &= \frac{1}{4\frac{g^2}{\gamma_s^2} v^2 + (w-v)^2} \left( 4\frac{g^2}{\gamma_s^2} v^2 \sqrt{u + w} + (w-v)^2 \sqrt{u-w} \right), \\
B &= \frac{1}{4\frac{g^2}{\gamma_s^2} v^2 + (w-v)^2} \left( (w-v)^2  \sqrt{u+w} + 4\frac{g^2}{\gamma_s^2} v^2 \sqrt{u - w}  \right), \\
C &= \frac{-2\frac{g}{\gamma_s} v \,(w-v)}{4\frac{g^2}{\gamma_s^2} v^2 + (w-v)^2}  \left(  \sqrt{u+w} - \sqrt{u-w}\right),
\end{align*}
in terms of the parameters $u,v,w$ defined as follows
\begin{align*}
u= \frac{c+d}{2}, \quad v=\frac{c-d}{2}, \quad w= \frac{|c-d|}{2}\sqrt{1+4\frac{g^2}{\gamma_s^2}}.
\end{align*}

The jump operators in the Doob dynamics read
\begin{align*}
& L_1^s= \mathrm{e}^{-s/2} \frac{\sqrt{\alpha}}{\sqrt{a}} \Big( A\ketbra{2}{1} -iC \ketbra{3}{1} \Big), \quad \quad \quad\,\, L_5^s=\mathrm{e}^{s/2}\frac{\sqrt{\alpha}}{\sqrt{a}}  \Big( B\ketbra{3}{1} +iC \ketbra{2}{1} \Big), \\
& L_2^s=\mathrm{e}^{-s/2}\frac{\sqrt{\alpha} \sqrt{a}}{AB-C^2}\Big( B\ketbra{1}{2} -iC \ketbra{1}{3} \Big), \quad  L_6^s=\mathrm{e}^{s/2}\frac{\sqrt{\alpha} \sqrt{a}}{AB-C^2}\Big( A\ketbra{1}{3} +iC \ketbra{1}{2} \Big) ,\\
& L_3^s=\mathrm{e}^{-s/2} \frac{\sqrt{\alpha}}{\sqrt{a}} \Big( A\ketbra{2}{4} -iC \ketbra{3}{4} \Big), \quad \quad \quad\,\, L_7^s=\mathrm{e}^{s/2}\frac{\sqrt{\alpha}}{\sqrt{a}}  \Big( B\ketbra{3}{4} +iC \ketbra{2}{4} \Big), \\
& L_4^s=\mathrm{e}^{-s/2}\frac{\sqrt{\alpha} \sqrt{a}}{AB-C^2}\Big( B\ketbra{4}{2} -iC \ketbra{4}{3} \Big), \quad L_8^s=\mathrm{e}^{s/2}\frac{\sqrt{\alpha} \sqrt{a}}{AB-C^2}\Big( A\ketbra{4}{3} +iC \ketbra{4}{2} \Big) .
\end{align*} 
Note that the symmetry under the exchange $2 \leftrightarrow 3$ is broken, but still we have the symmetry under the switch $1 \leftrightarrow 4$.

\end{widetext}

\section{SCGF in the third example}
\label{app: ex3}

First of all, one can notice that in each block labeled by $d$, the Hamiltonian part of the generator has still a finer block diagonal structure. Indeed, for instance, in $\mathcal{L}_\lambda^1$ with just one matrix $\sigma_y$ or $\sigma_x$ the Hamiltonian can only connect those basis elements with the $\sigma_{x/y}$ on the same position and equal matrices on the rest of the chain apart from the $x/y$ nearest neighbours. Therefore, one has for any position $j$ in the chain and for any choice of the $N-3$ sites that exclude $j,j+1,j-1$, an eight-dimensional sub-block spanned by $\ket{\ldots zxz \ldots}, \ket{\ldots 1xz \ldots}, \ket{\ldots zx1 \ldots}, \ket{\ldots 1x1 \ldots},$ $ \ket{\ldots zyz \ldots}, \ket{\ldots 1yz \ldots}, \ket{\ldots zy1 \ldots}, \ket{\ldots 1y1 \ldots}$,
where the Hamiltonian part of the generator has the following antisymmetric representation (taking the basis vectors in the order mentioned previously)
\begin{equation}
\begin{pmatrix}
0 &0 &0 &0 &0 &2J &2J &0 \\
0 &0 &0 &0 &2J &0 &0 &2J \\
0 &0 &0 &0 &2J &0 &0 &2J \\
0 &0 &0 &0 &0 &2J &2J &0 \\
0 &-2J &-2J &0 &0 &0 &0 &0\\
-2J &0 &0 &-2J &0 &0 &0 &0\\
-2J &0 &0 &-2J &0 &0 &0 &0\\
0 &-2J &-2J &0 &0 &0 &0 &0
\end{pmatrix}.
\end{equation}
More in general, for any sub-block, in any $d$-block, one can arrange the basis states in such a way that the Hamiltonian part of the generator is antisymmetric. This is because the Hamiltonian always switches $\sigma_x$ to $\sigma_y$ with a coefficient $-2J$ and $\sigma_y$ to $\sigma_x$ with a coefficient $2J$. This structure is very important for our purposes.

Another important thing to notice is that, at least concerning the diagonal elements, coming from the dissipative part of the generator, the $\theta(\lambda)$ presented in the main text is indeed the largest value (all these entries are real). Therefore, by subtracting in each block $B$ the term $\theta(\lambda)\mathbbm{1}$ one can rephrase the original problem into the following: show that the eigenvalues of $B-\theta(\lambda)\mathbbm{1}$ have negative real part. This is indeed the case due to the theorem below. 
\\

\textbf{Theorem} Consider two $n \times n$ real matrices $A$ and $D$, such that $A=-A^T$ and $D= (D_j \delta_{ij})_{ij}$, with $D_j<0$. Then the eigenvalues of $A+D$ have negative real part. 
\begin{proof}
Consider an eigenvalue $\lambda$ of $A+D$, with eigenvector $\underline{v}$ normalized to $1$, namely 
\begin{equation}
(A+D) \cdot \underline{v} = \lambda \underline{v} , \quad (\underline{v}^* )^T  \cdot \underline{v}= 1.
\end{equation}
Therefore one has
\begin{equation}\label{proof1}
(\underline{v}^* )^T (A+D) \cdot \underline{v} = \lambda, 
\end{equation}
and taking the conjugate transpose of the previous equation one also has
\begin{equation}\label{proof2}
(\underline{v}^* )^T (A^T+D) \cdot \underline{v} = \lambda^*. 
\end{equation}
Summing \eqref{proof1} and \eqref{proof2}, because $A$ is antisymmetric one finds
\begin{align}
\mathrm{Re}(\lambda) = (\underline{v}^* )^T D\cdot \underline{v} = \sum_j |v_j|^2 D_j <0,
\end{align}
that is the thesis we wanted to prove.
\end{proof}

\bibliography{referencias-BibDesk-OK}{}

\end{document}